# Optical capabilities of the Multichannel Subtractive Double Pass (MSDP) for imaging spectroscopy and polarimetry at the Meudon Solar Tower

*Malherbe J.-M. (1), Mein P. (1), Sayède F. (2)*
*Observatoire de Paris, Section de Meudon, LESIA (1), GEPI (2), 92195 Meudon, France*

*November 26, 2023*

## Abstract

The Meudon Solar Tower (MST) is a 0.60 m telescope dedicated to spectroscopic observations of solar regions. It includes a 14-meter focal length spectrograph which offers high spectral resolution. The spectrograph works either in classical thin slit mode (R > 300000) or 2D imaging spectroscopy (60000 < R < 180000). This specific mode is able to provide high temporal resolution measurements (1 min) of velocities and magnetic fields upon a 2D field of view, using the Multichannel Subtractive Double Pass (MSDP) system. The purpose of this paper is to describe the capabilities of the MSDP at MST with available slicers for broad and thin lines. The goal is to produce multichannel spectra-images, from which **cubes** of **instantaneous data (x, y, λ)** are derived, in order to study of the plasma dynamics and magnetic fields (with polarimetry).

## Introduction

The Multichannel Subtractive Double Pass (MSDP) is an imaging spectroscopy technique introduced by Mein (1977) at the Meudon Solar Tower (MST). It is based on a slicer which provides line profiles with N sampling points (or N channels) over a 2D field of view (FOV); for that purpose, the MSDP has a rectangular entrance window instead of a thin slit. The technique was progressively developed and implemented on many telescopes (Mein *et al*, 2021). The first instrument with N = 7 (later 9) channels was incorporated to the 14-meter spectrograph of MST. It was mainly working with the Hα line in order to study the dynamics of chromospheric features. It was soon followed by the MSDP of the Pic du Midi Turret Dome (Mein, 1980) with N = 11 channels. The third instrument was integrated to the 15-meter spectrograph of the german Vacuum Tower Telescope (VTT) in Tenerife (Mein, 1991). Meanwhile, polish colleagues introduced the technique on the large Bialkow coronagraph to observe prominences (Rompolt, 1994). The last instrument was designed for the 8-meter spectrograph of the THEMIS telescope (Mein, 2002) with N = 16 channels.

The first generation slicers were based on multi-slits in the spectrum and prism beam-shifters. However, this technique does not allow to increase much the spectral resolution (limited to about 80 mÅ) and the number of channels. For that reason, the second generation of slicers uses now micro-mirrors, which allow to reach about 30 mÅ of spectral resolution and could deliver up to 50 channels. The first slicer using this technology was installed ten years ago at MST (N = 18 channels). The Solar Line Emission Dopplerometer for the Lomnicky coronagraph (N = 24 channels) is a project (France, Poland, United Kingdom, Slovakia) dedicated to velocity measurements in the hot corona (Malherbe *et al*, 2021). The present paper discusses the capabilities of the available slicers at MST for imaging spectroscopy of broad chromospheric lines (100-300 mÅ spectral resolution, N = 9 channels) and thin photospheric lines (30 mÅ resolution, N = 18 channels). The MSDP can deliver maps of dopplershifts at several altitudes, as shown by figure 1. A special polarimeter is able to split each channel into two sub-channels for measurements of simultaneous Stokes combinations (I+S and I-S, where S = Q, U, V in sequence), providing the possibility to derive maps of magnetic fields.

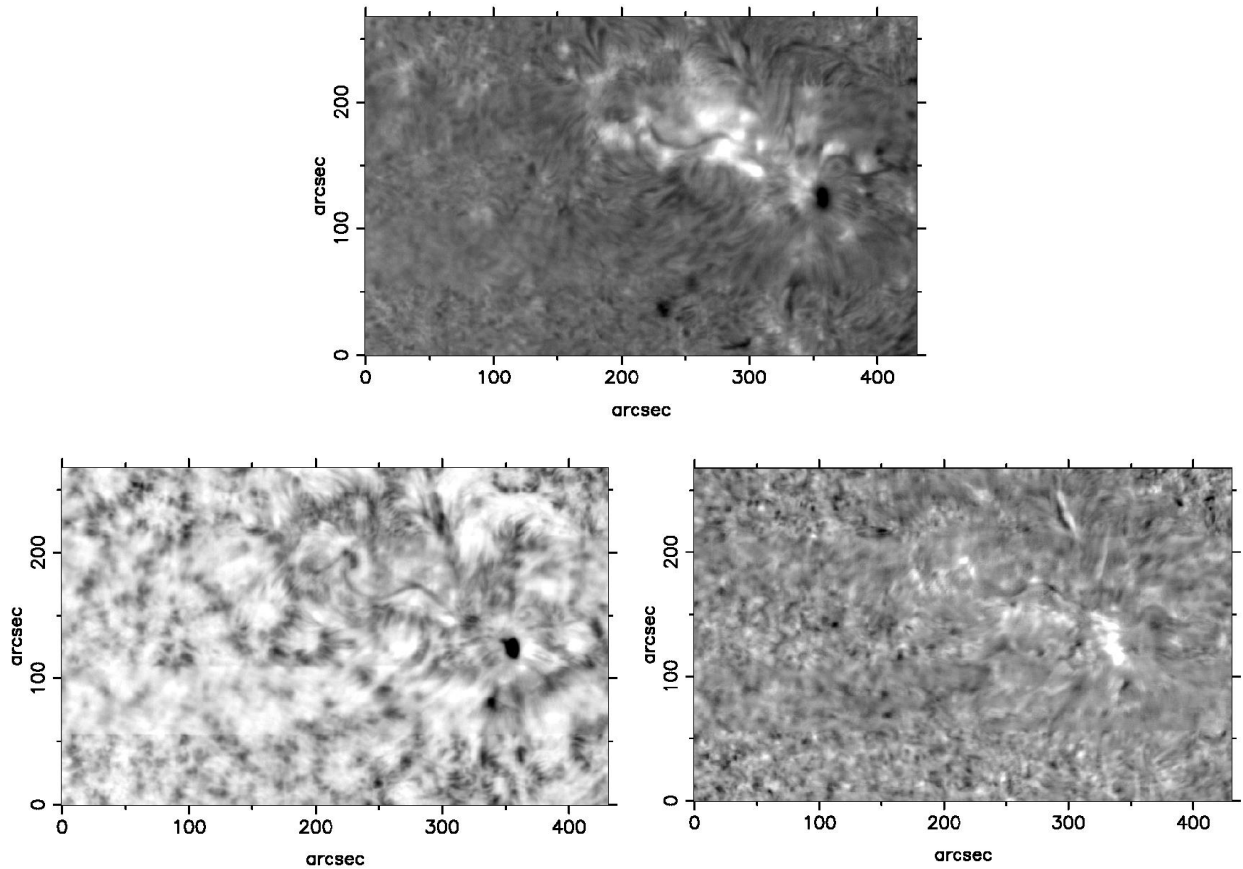

**Figure 1:** *MSDP FOV at two altitudes in Hα. Top: line centre. Bottom: inflexion points ±0.45 Å apart from line core; from left to right, intensity fluctuations and dopplershifts (grey levels). Courtesy Paris Observatory.*

## 1 - Characteristics of the telescope and spectrograph

The MST is a 0.60 m aperture/45 m focal length vertical telescope fed by a coelostat (two 0.80 m flat mirrors). With the optional focal reducer, the focal length becomes 22.5 m and a 30 cm diaphragm is used in order to keep the same focal/aperture ratio (F/75).

A large horizontal spectrograph is fed by the telescope. The collimator and camera mirrors both have 14 m focal length. The grating has 300 grooves/mm and b = 63°26' is the blaze angle. In the blaze (i = i' = b), we have n λ = 59628 Å, where n is the interference order. Fraunhofer lines such as CaII K/H, NaD1, Hα or the CaII infrared triplet, appear, respectively, in orders n = 15, 10, 9 or 7. Order selection requires the use of 100 Å bandwidth interference filters. The dispersion is d = n x 0.94 mm/Å, so that typical values lie in the range d = 6-14 mm/Å and depend on the observed line (better dispersion towards the blue). With a thin slit of width 0.15 mm, we have R = 370000 and the spectral resolution lies around 20 mÅ in the red part of the spectrum (13 mÅ in the blue); this 0.15 mm slit corresponds to 0.7" on the Sun for the 45 m focal length of the telescope (1.4" with the reducer). A CCD array (1370 x 1040 pixels) is available. For imaging spectroscopy, the slit is replaced by a rectangular window (or field stop), for instance of 6.5 mm x 52 mm with standard slicers; the resulting FOV is 30" x 4' for 45 m focal length, or 1' x 8' with the 22.5 m reducer. The slicer is placed in the spectrum; it is a beam splitter-shifter which produces N cospatial channels (figure 2). Each one is characterized by the number of channels (N), the step in the spectrum between two consecutive channels (Δx in mm or Δx/d in Å) and the shift w (in mm) between consecutive channels.

Two slicers are available for broad lines; their characteristics are N = 9 channels,  Δx = 2.5 mm (for the broad Hα line) or Δx = 1.0 mm (for lines such as NaD1 or the infrared triplet of CaII), and  w = 9.0 mm (in practice 6.5 mm are usable, this is the width of the associated entrance window). These slicers are displayed in figure 3a. They use a multi-slit in the spectrum (beam splitting) and prisms as beam shifters. The width of the MSDP image provided by these slicers in the spectrum is N x w = 9 x 9 = 81 mm. Many sequences were recorded on 70 mm films, until the integration of a CCD in the early 2000s.

There is also a new generation slicer at MST with N = 18 channels, Δx = 0.3 mm, designed for thin photospheric lines, and w = 3.3 mm (this is the width of the corresponding entrance window of 15" x 2' with 45 m focal length, or 30" x 4' with the 22.5 m focal reducer). In this slicer (figure 3b), the multi-slit was replaced by manufactured micro-mirrors (beam splitter), while beam shifting is no more done by prisms, but by adjustable small mirrors. The spectral resolutions of 30 mÅ can be achieved, so that observations of thin lines of FeI, which are mostly sensitive to phostospheric magnetic fields, become quite possible. The width of the MSDP image provided by this slicer in the spectrum is N x w = 18 x 3.3 = 60 mm.

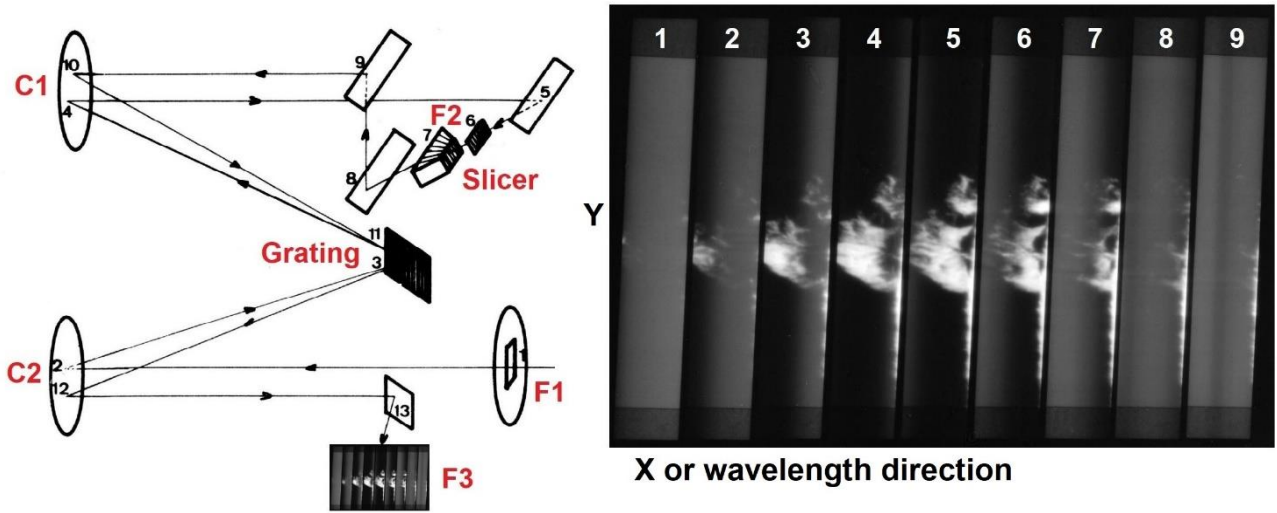

*Figure 2:* MSDP optical principle at MST. F1 = entrance window. C1 = camera mirror. C2 = collimator. F2 = spectrum with slicer. F3 = output (Hα spectra-image with N = 9 channels). The focal length is 14 m (courtesy Paris Observatory).

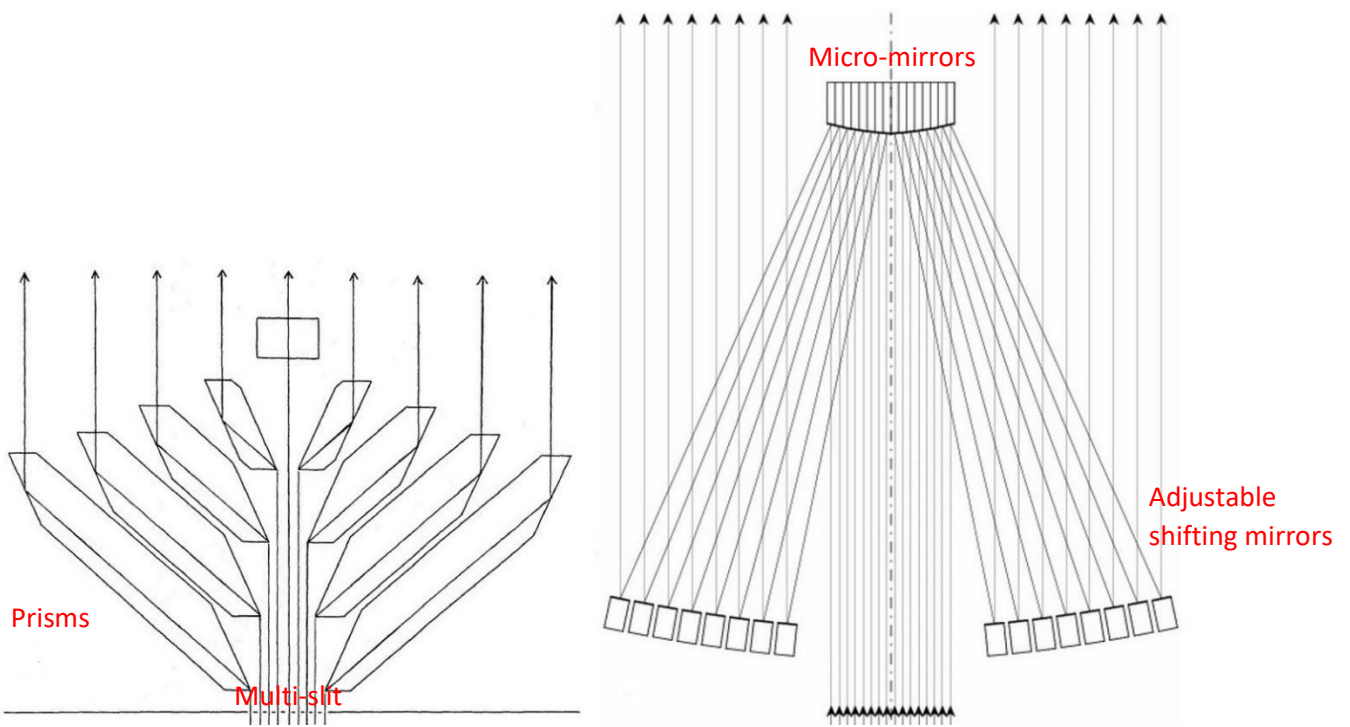

*Figure 3:* Left (a) : 9-channel slicer with multi-slit splitter and prism beam shifer. Right (b) : 16-channel slicer with micro-mirrors for beam splitting and adjustable mirrors for beam shifting (courtesy Paris Observatory).

Figure 4 shows the implementation of the new generation 18-channel slicer in the MST spectrograph. Figure 5 details the light path inside the slicer. The beam splitting is done by manufactured micro-mirrors (0.3 mm width) with no possible adjustment (mono block device). On the contrary, beam shifting is achieved by small individual mirrors, which are individually adjustable.

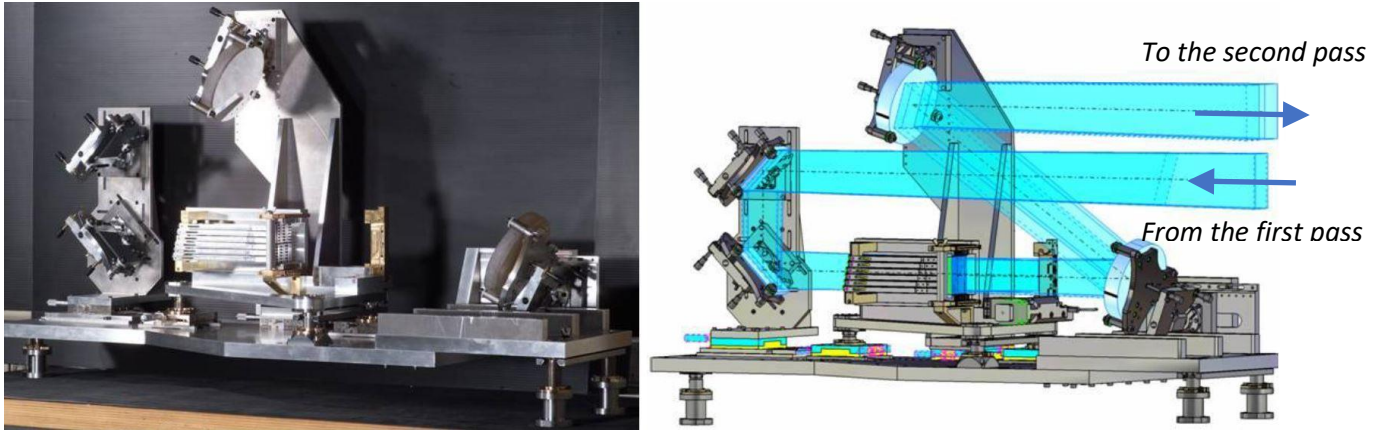

*Figure 4:* *The slicer in the spectrum at MST and the 4 folding mirrors around (courtesy Paris Observatory).*

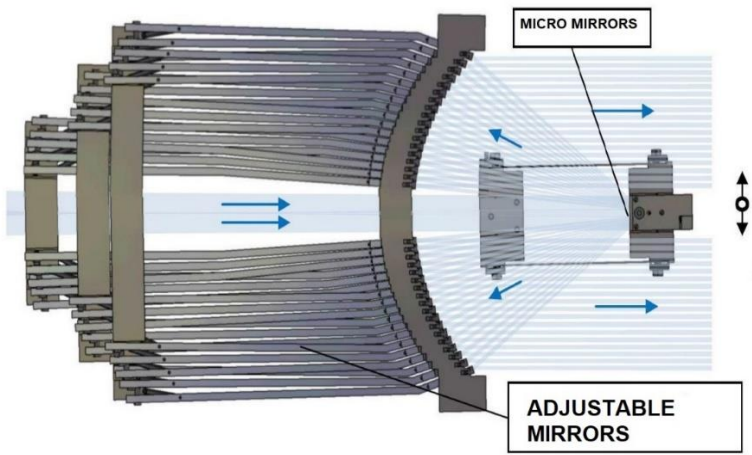

*Figure 5:* *details of the new generation slicer with micro mirrors (mono block beam splitter) and adjustable mirrors for beam shifting (courtesy Paris Observatory).*

## 3 – Sampling of broad lines with the 9-channel slicers

Figures 6, 7 & 8 display the wavelength transmission of the 9-channel slicers, respectively for Hα, NaD1 and CaII 8542 spectral lines. The figures indicate the atlas profiles at disk centre for reference, together with the first derivative: it allows to locate the inflexion points, which play a major role for the determination of LOS magnetic fields in the frame of the weak field theory (the Stokes V profile is proportional to the line derivative and reaches maximum values at the inflexion points). The wavelength functions are given by:

$$\lambda_n(x) = \lambda_0 + (x/d) + n\,(\Delta x/d)$$

for $0 < x < x_m$ ($x_m = 6.5$ mm is the maximum FOV in x-direction). In this formula, n is the current channel ($1 < n < N$, $N = 9$), d is the dispersion of the spectrograph (mm/Å), $\Delta x$ the slicer step (mm, the spectral resolution in Å is $\Delta x/d$).

Two slicers are available with $\Delta x = 2.5$ mm (figure 6 for broad lines such as Hα) and $\Delta x = 1.0$ mm for (figures 7 & 8 for Fraunhofer lines except Hα). The values (mm) of the abscissa (x) correspond to the following scale: 1 mm = 4.6" or 9.2", respectively for the 45 m (standard) or 22.5 m (reduced) focal length of the telescope. $x_m = 6.5$ mm provides the largest FOV in x-direction, 30" or 1' according to the focal length. In y-direction, the FOV is 8 times larger, 4' or 8'.

Hα is a very broad line (1.0 Å FWIP, Full Width at Inflexion Points) which requires a large step slicer (2.5 mm or 0.30 Å). Assuming a reasonable bandwidth for the profile (1.5 Å), line of sight (LOS) velocities up to 10 km/s can be measured over a 17" wide FOV (figure 6), more with the focal reducer of the telescope (34").

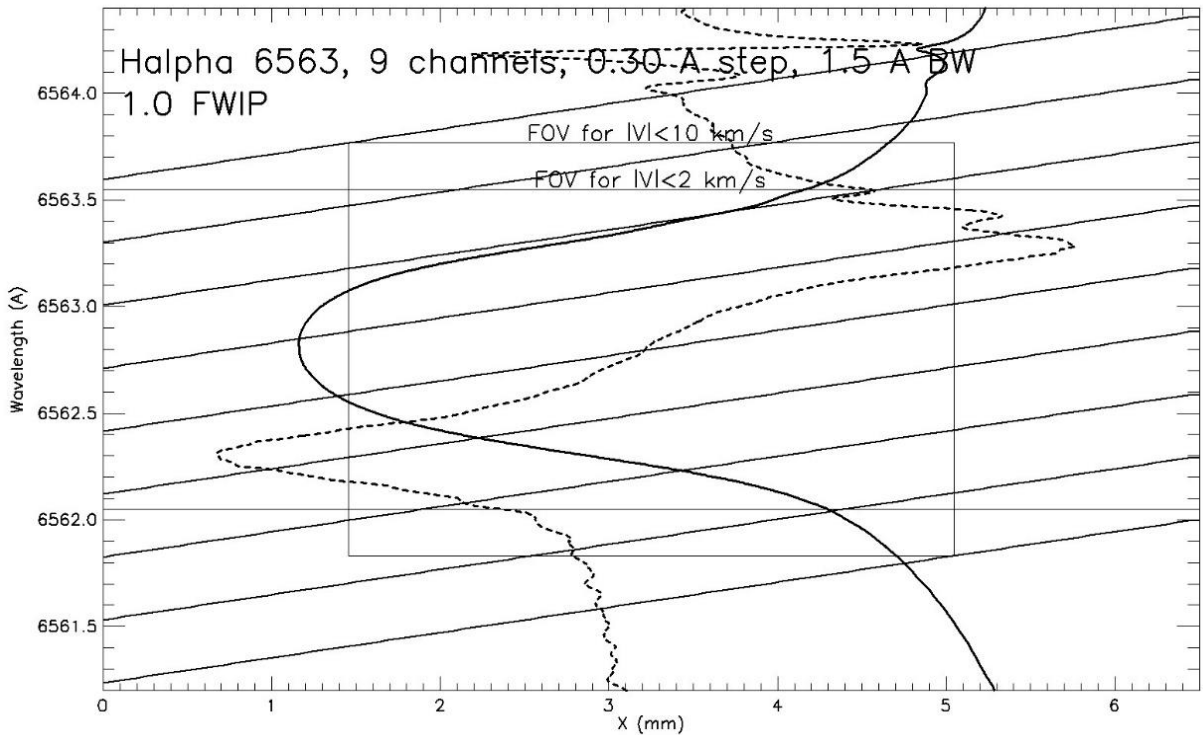

**Figure 6:** *the wavelength transmission (Å) of the 9-channel slicer for the Hα line as a function of the abscissa (mm). The atlas profile at disk centre is shown, together with its first derivative (courtesy Paris Observatory).*

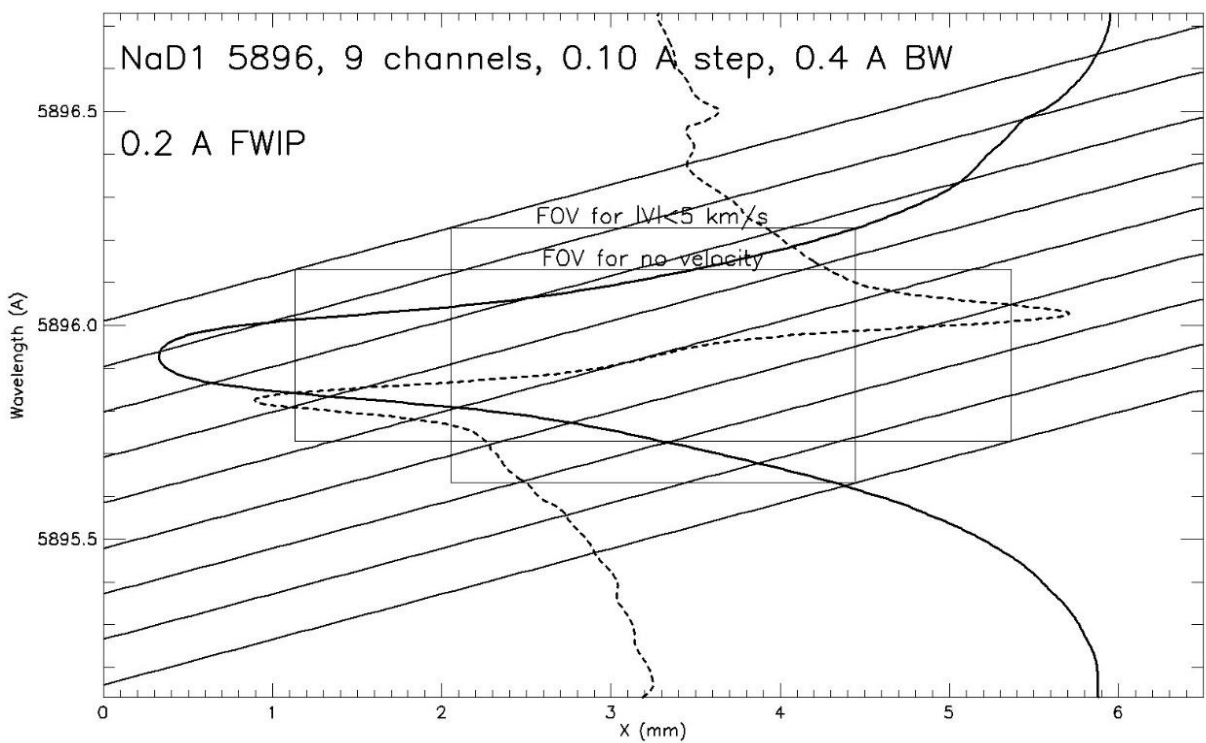

**Figure 7:** *the wavelength transmission (Å) of the 9-channel slicer for the NaD1 line as a function of the abscissa (mm). The atlas profile at disk centre is shown, together with its first derivative (courtesy Paris Observatory).*

NaD1 is 5 times thinner than Hα (0.2 Å FWIP), which requires a smaller step slicer (1.0 mm or 0.10 Å). Assuming a given 0.4 Å bandpass for the profile, LOS velocities up to 5 km/s can be measured over a 11" wide FOV (figure 7), more with the focal reducer of the telescope (22"). The same slicer allows to observe CaII 8542 (0.4 Å FWIP) with 0.15 Å resolution (figure 8), but this line is two times broader than NaD1, so that the number of channels limits the size of the FOV in the x-direction when velocities are large.

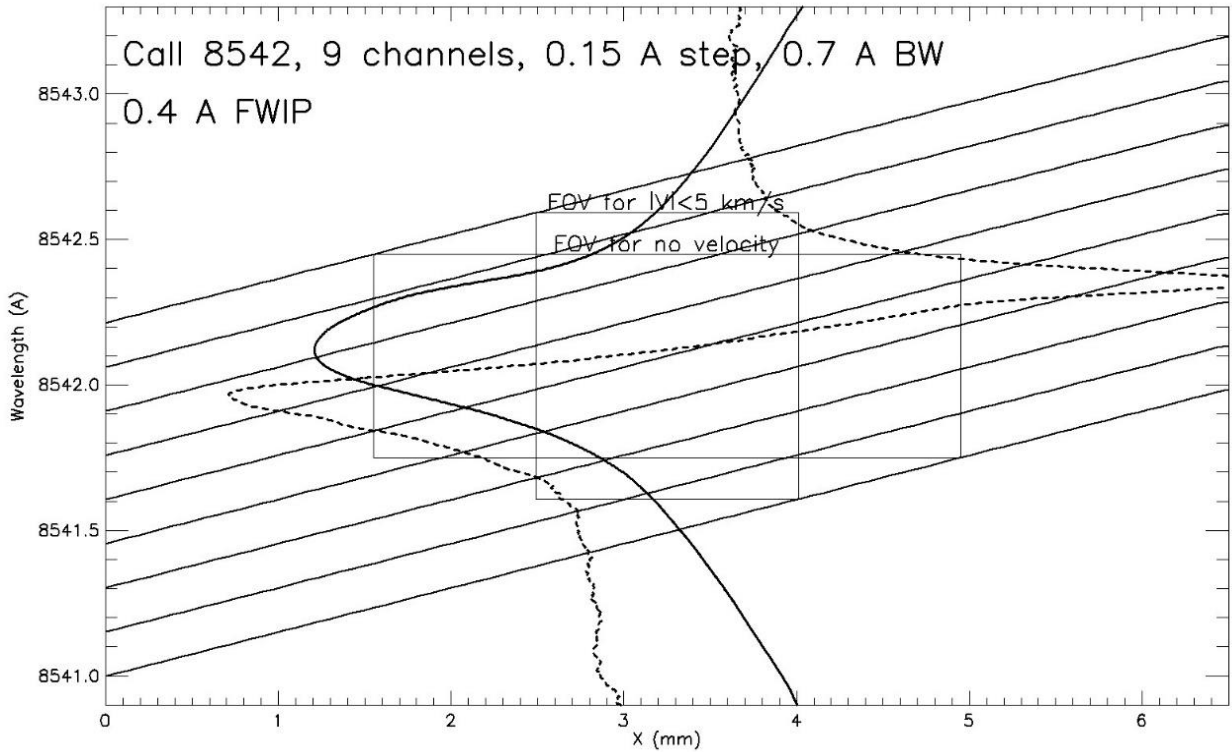

***Figure 8:*** *the wavelength transmission (Å) of the 9-channel slicer for the CaII 8542 line as a function of the abscissa (mm). The atlas profile at disk centre is shown, together with its first derivative (courtesy Paris Observatory).*

### 4 – Sampling of thin lines with the 18-channel slicer

Figures 9 & 10 display the wavelength transmission of the 18-channel slicer, respectively for Mg b2 5173 and FeI 6173. These lines are too thin to be studied with the 9-channel slicers, which deliver, in the best case, 0.10 Å spectral resolution. Photospheric lines require better than 30 mÅ. The FOV in the x-direction is 3.3 mm maximum (corresponding to 15" or 30", respectively for 45 m or 22.5 m focal length of the telescope). It is 8 times larger in the y-direction.

Mgb2 has 0.15 Å FWIP. The spectral resolution provided by the slicer for this line is 29 mÅ. This is an excellent result (figure 9), but the necessary bandpass of 0.25 Å around the line core does not allow, with the 18-channel slicer, to observe a wide FOV when large dopplershifts are present.

Observations of the thin photospheric FeI 6173 line provide a larger FOV in the case of dopplershifts (figure 10), because the width of the line (0.085 Å) does not require to record a large bandwidth around the line core (0.16 Å is sufficient). For this bandpass, with dopplershifts of ±5 km/s, the x-direction FOV is reduced by a factor 2, which is acceptable. The spectral resolution of the slicer is 32 mÅ for that line. FeI 6173 is close to the gaussian shape and is observed by many instruments which do not have such a resolution, in order to derive magnetic fields (HMI onboard the Solar Dynamics Observatory SDO/NASA or the PHI onboard the ESA Solar Orbiter mission). The 18-channel slicer of MST avoids shape assumptions and is therefore convenient for most photospheric lines.

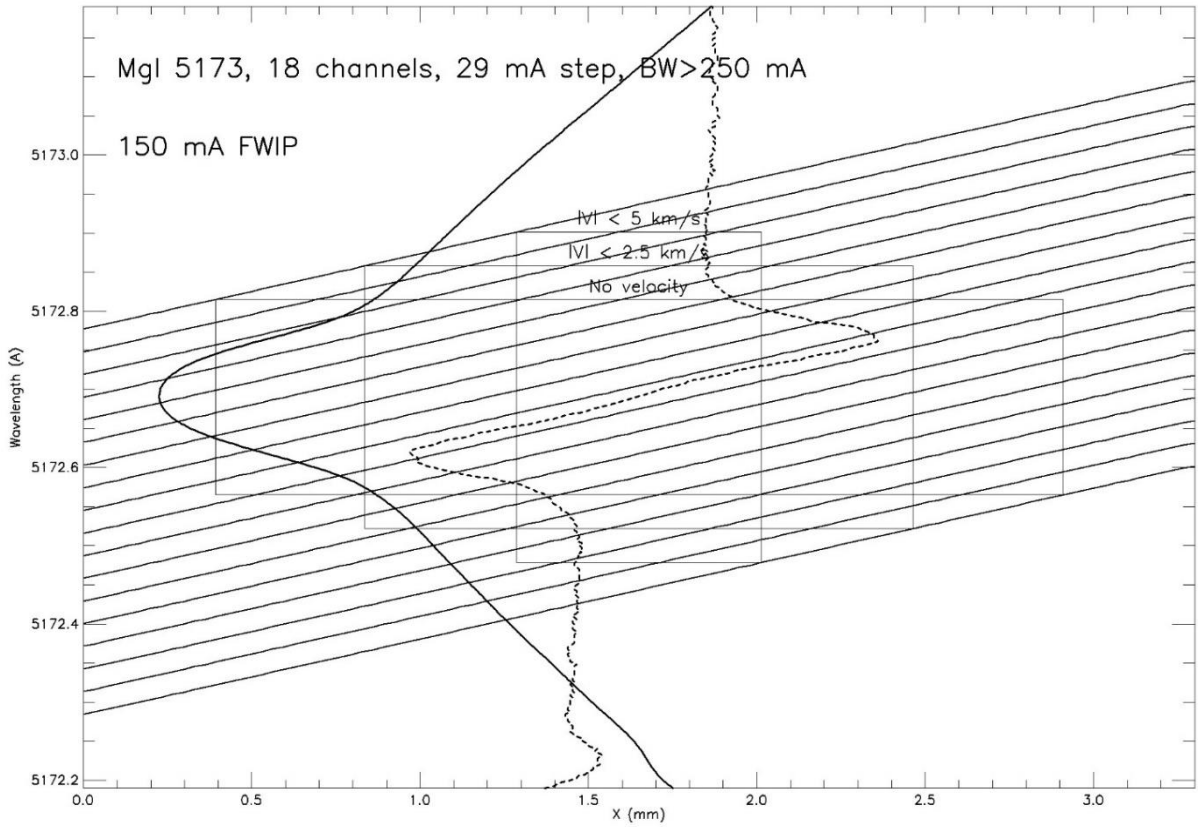

*Figure 9: the wavelength transmission (Å) of the 18-channel slicer for the Mg b2 5173 line as a function of the abscissa (mm). The atlas profile and its first derivative are drawn (courtesy Paris Observatory).*

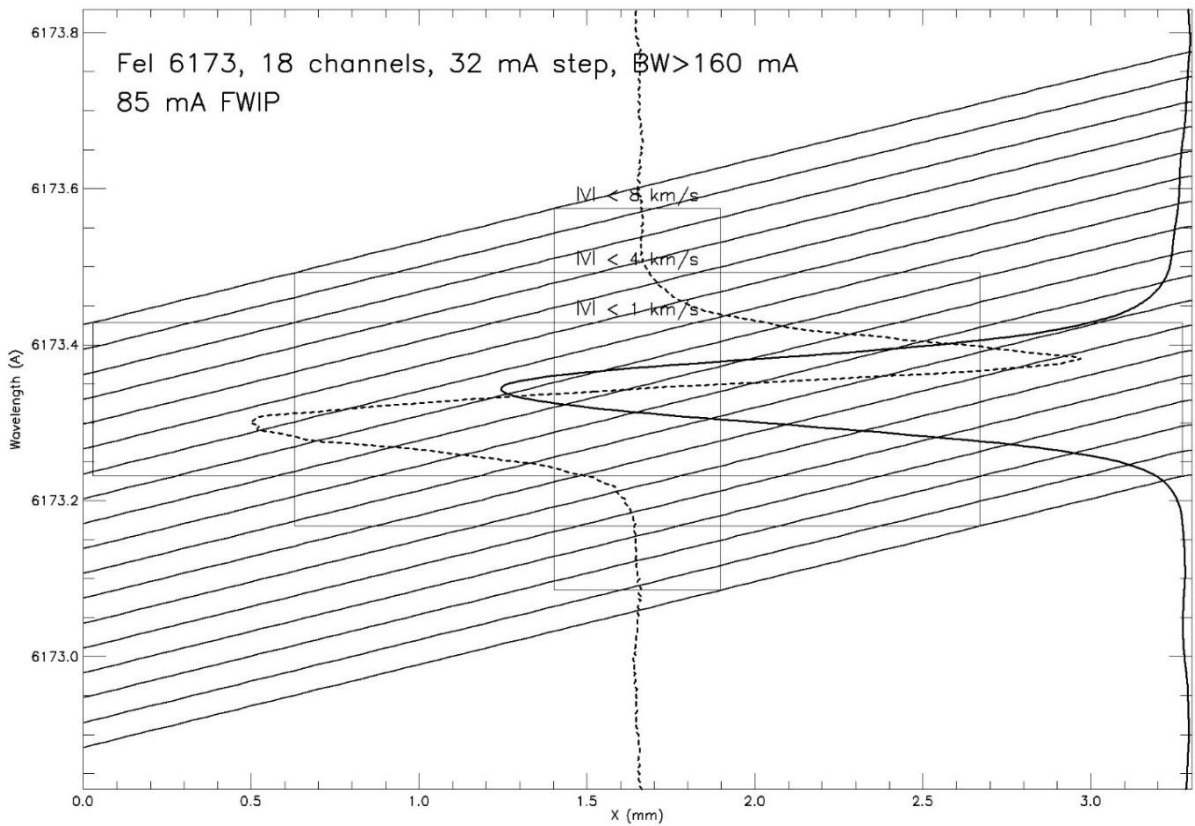

*Figure 10: the wavelength transmission (Å) of the 18-channel slicer for the FeI 6173 line as a function of the abscissa (mm). The atlas profile and its first derivative are drawn (courtesy Paris Observatory).*

## 5 – MSDP polarimetry with 2 x 18 sub-channels

As the seeing at MST is never better than 1.5''-2'', the sequential observation of Stokes combinations, such as I+Q, I-Q, I+U, I-U, I+V, I-V is not recommended. Moreover, four flat mirrors are present in the light path: the two mirrors of the coelostat (with varying incidence) and two other mirrors (with fixed incidence) to redirect the light beam towards the spectrograph. Hence, there is instrumental polarization, so that I+V and I-V only (circular polarization) can be observed with confidence. The measures must be preferably simultaneous because of uncorrected turbulence. For that purpose, we use a dual beam method, not far from the one introduced by Semel (1980). It consists in splitting each MSDP channel in two sub-channels of half width, providing cospatial signals I+V and I-V. The wavelength transmission is now provided by the law:

Sub-channel I-V : $\lambda_n(x) = \lambda_0 + (x/d) + n\,(\Delta x/d)$

Cospatial sub-channel I+V : $\lambda_n(x) = \lambda_0 + [\,(x + x_m/2)\,/d\,] + n\,(\Delta x/d)$

for $0 < x < x_m/2$ ($x_m/2 = 1.65$ mm is now the maximum FOV in x-direction, it corresponds either to 7.5'' or 15'', respectively for the telescope focal length of either 45 m or 22.5 m). These functions show that the wavelength sampling of I+V and I-V profiles is not identical, so that interpolations must be performed in order to combine them and derive the Stokes profiles of $I(\lambda)$, $V(\lambda)$ and the polarization rate profile $V/I(\lambda)$.

In figure 11, the entrance window of the spectrograph ($0 < x < x_m$), at the focus of the telescope, is replaced by a rectangular field stop of half width ($0 < x < x_m/2$) followed by a birefringent beam splitter-shifter (calcite). It generates a dual beam with orthogonal linear polarizations, which are injected into the spectrograph. The two beams have a cospatial FOV and contain respectively I+V and I-V. The MSDP forms for each channel two sub-channels (with different sampling wavelengths). Hence, the output spectra-image is composed of 2 x N sub-channels provided by the N-channel slicer. In practice, N = 18, so that 36 sub-channels are delivered for the same FOV in two states of polarization.

We show below results of observations and instrumental simulations for Mgb2 5173 and FeI 6173 lines, which are sensitive, respectively, to magnetic fields of the low chromosphere and the photosphere.

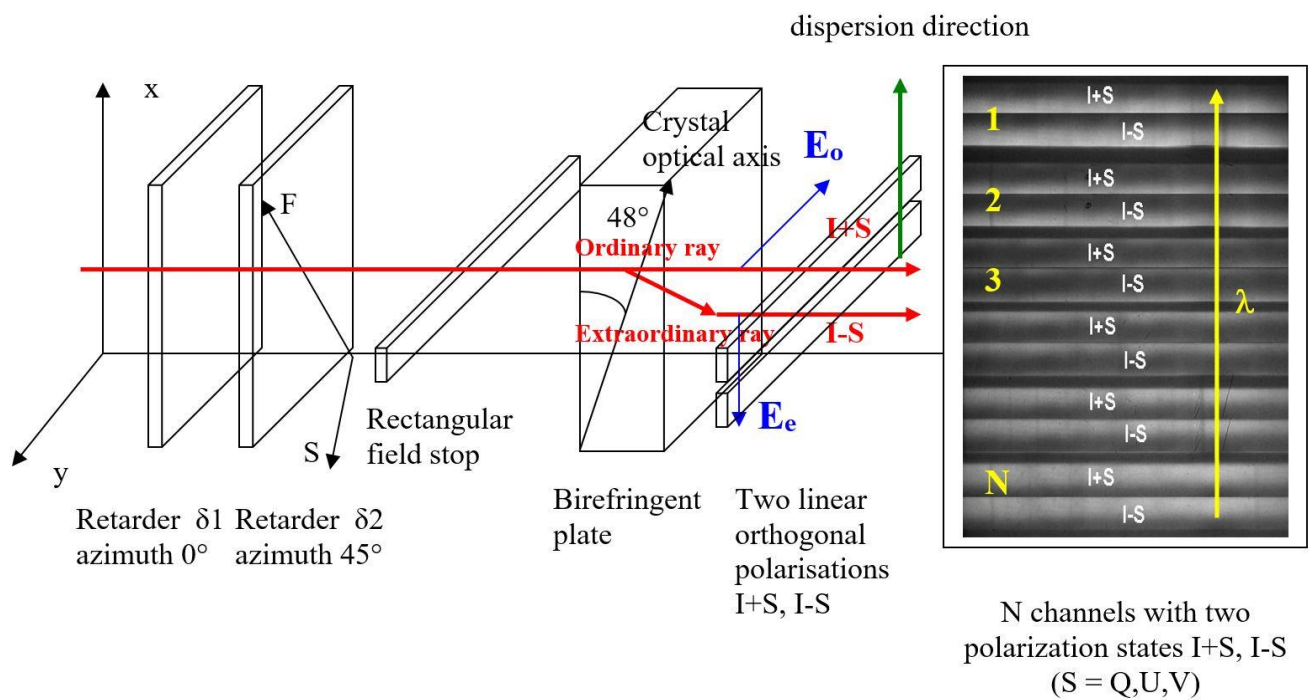

*Figure 11:* the dual beam polarimetric system used with the MSDP at the entrance of the spectrograph. One quarter-wave plate (azimuth 45°) allows to analyze the circular polarization. The linear polarization is, in theory, also observable with this device, using another waveplate (courtesy Paris Observatory).

### 5 – a – Simulation and observations of MgI b2 5173 (Landé factor g* = 1.75)

Figure 12 shows the wavelength functions of the sub-channels for Mg b2 (0.15 Å FWIP, 0.25 Å chosen bandwidth around the line centre, 29 mÅ spectral resolution). The left and right parts are cospatial, but the spectral line inside is not sampled by the same wavelength values, because, for a given x-abscissa of the FOV of width ($x_m$/2), there is a wavelength shift of $x_m$/(2d) between I+S and I-S. It appears in figures 13, 14 & 15.

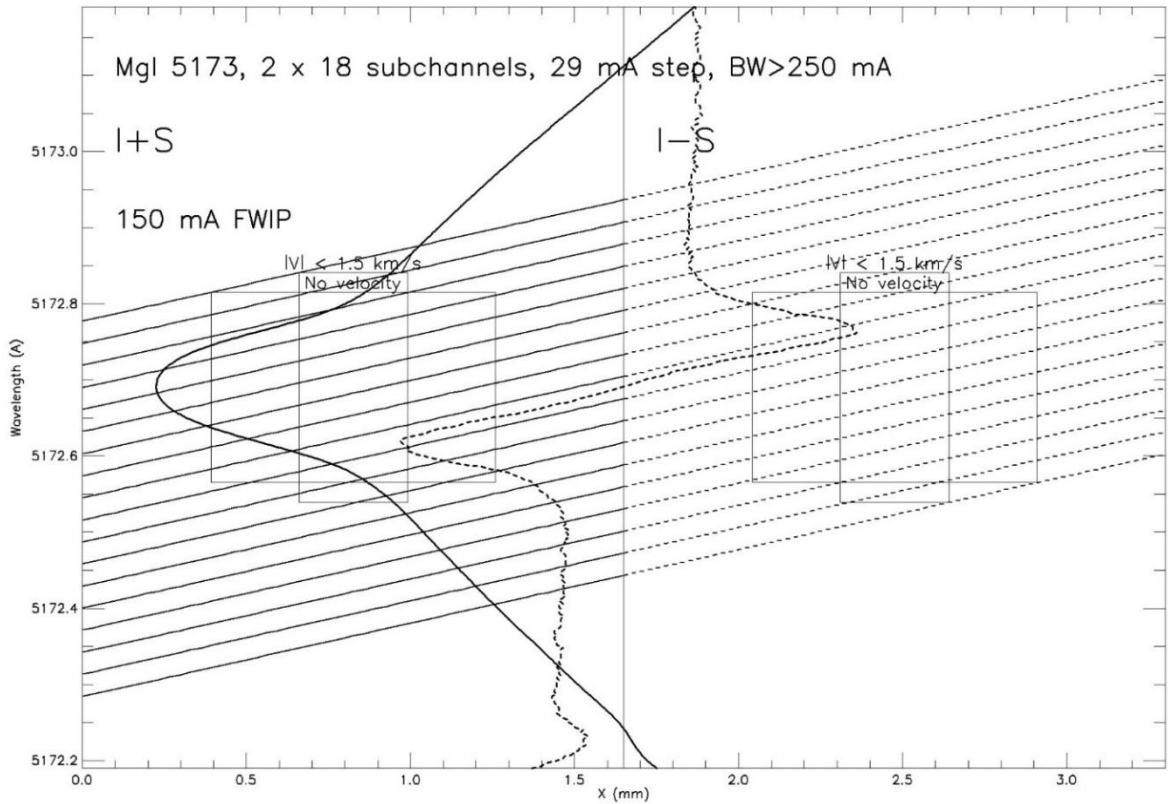

**Figure 12** : *wavelength functions of the 36 sub-channels for Mg b2 (courtesy Paris Observatory).*

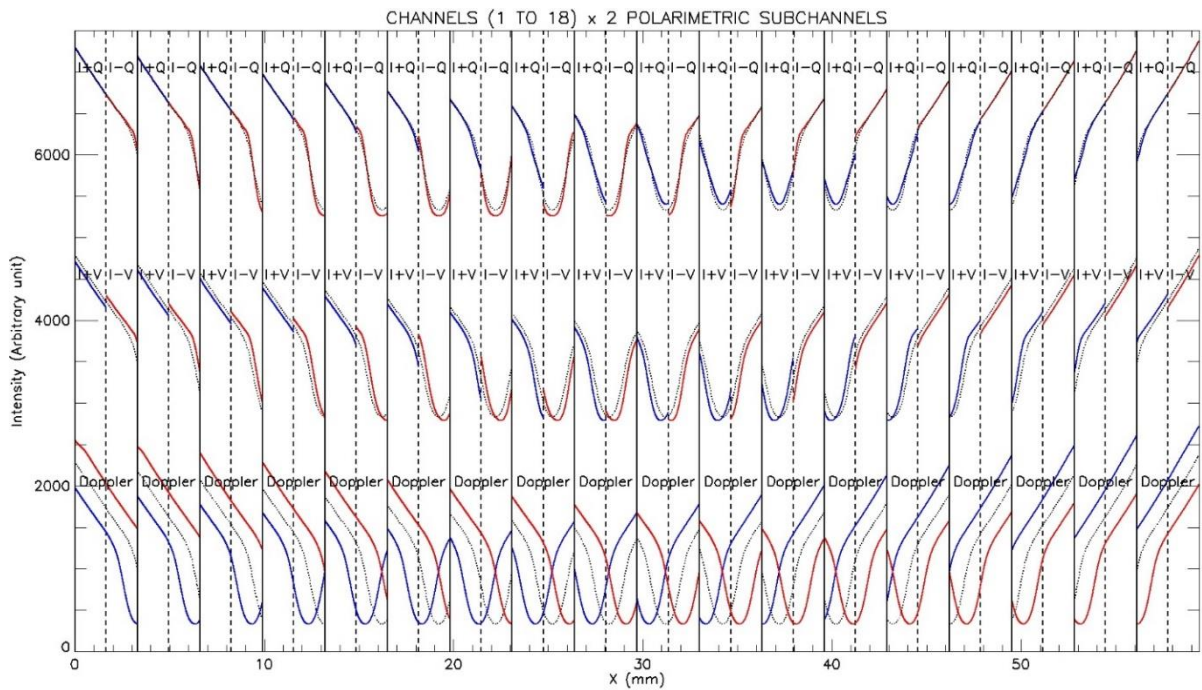

**Figure 13** : *Mg b2 spectral line in the 2 x 18 sub-channels. Top: I+Q and I-Q. Centre: I+V and I-V. Both are computed for B = 1000 G. Bottom: positive and negative dopplershifts (courtesy Paris Observatory).*

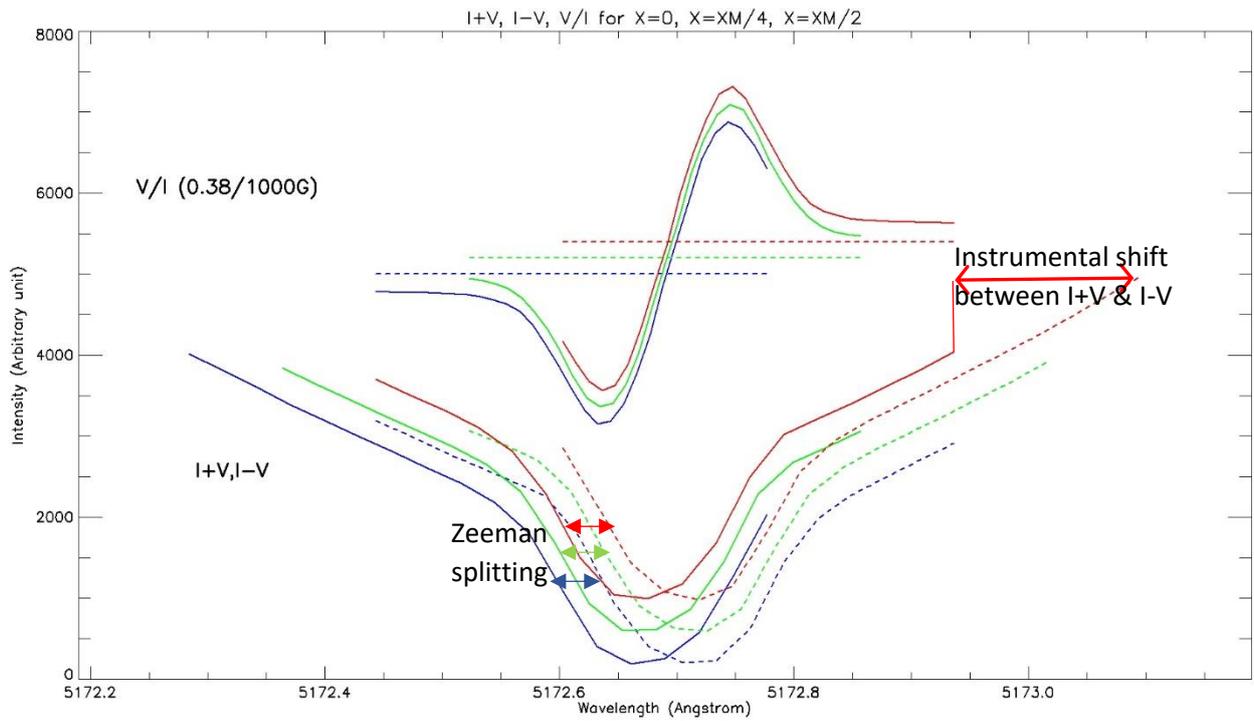

*Figure 14* : simulation of I+V (solid), I-V (dashed) and V/I profiles of Mg b2 spectral line for x = 0 (*blue*), $x_m/4$ (*green*) and $x_m/2$ (*red*). As the sampling wavelengths of I+V and I-V are not identical, an interpolation is necessary to get I, V and V/I. We assumed here B// = 1000 G (courtesy Paris Observatory).

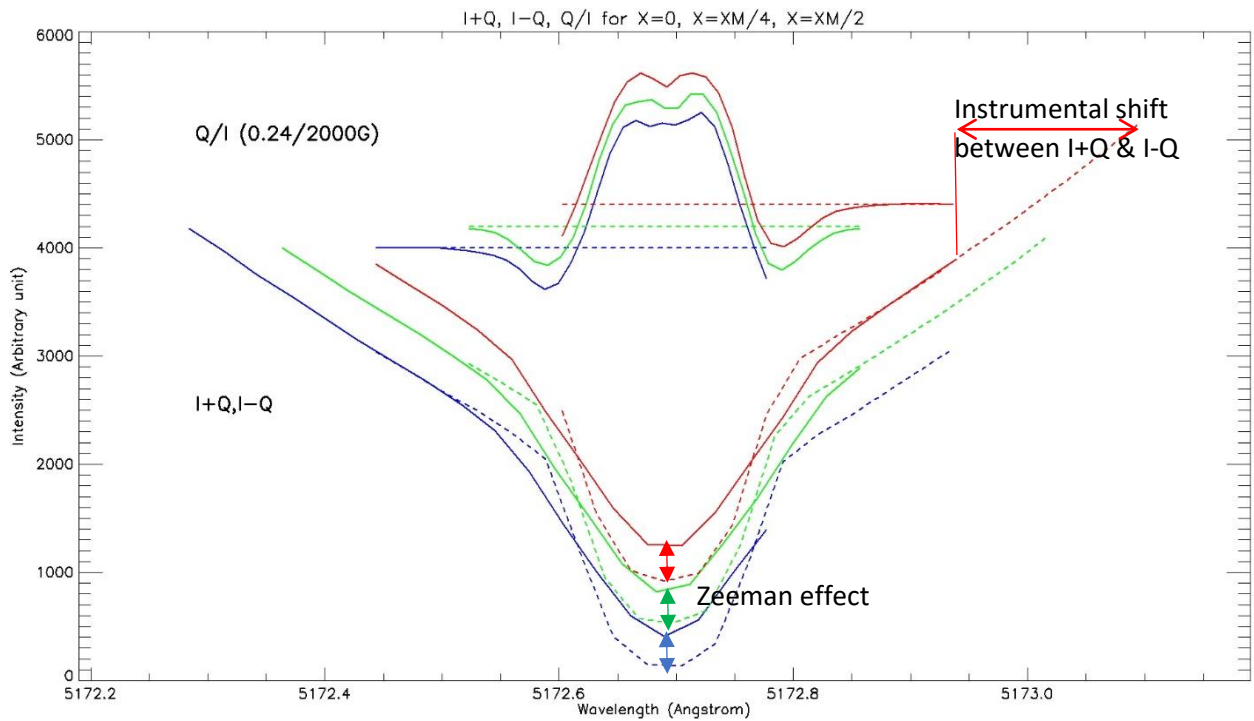

*Figure 15* : simulation of I+Q (solid), I-Q (dashed) and Q/I profiles of Mg b2 spectral line for x = 0 (*blue*), $x_m/4$ (*green*) and $x_m/2$ (*red*). As the sampling wavelengths of I+Q and I-Q are not identical, an interpolation is necessary to get I, Q and Q/I. We assumed B⊥ = 1000 G (courtesy Paris Observatory).

At last, figures 16 & 17 show an observation and the result after the data processing, in terms of intensity fluctuations, LOS velocities and LOS magnetic fields. Unfortunately, the seeing was poor, as it is frequent in Meudon. The final FOV of 170" x 300" is the juxtaposition of 25 individual observations of 15" x 170" obtained with an optical mechanical device, scanning the solar surface by 25 steps of 12" (total 300").

The scan takes 3 minutes, and one has to do also a Flat Field for channel detection (made at disk centre) and a dark current. It could be must faster with an optimized translator and a modern CCD.

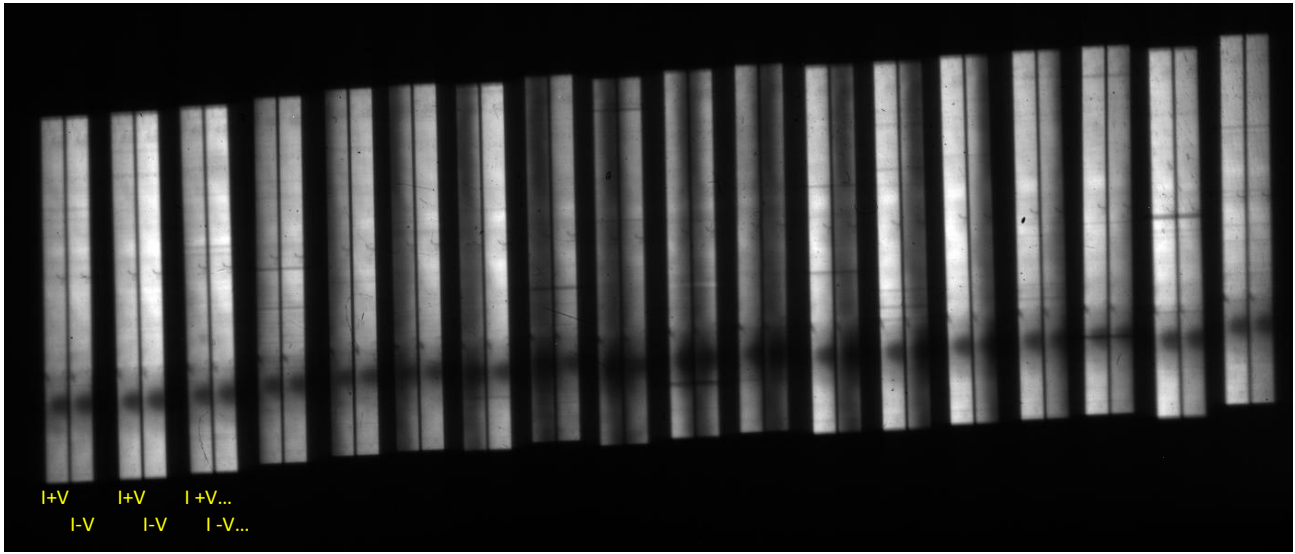

*Figure 16 : observation of an active region with the MSDP in Mg b2 line. The slicer and the polarimeter deliver 2 x 18 sub-channels for two states of circular polarization (I+V, I-V). 15" x 170" individual FOV. An active region requires a scan with at least 25 steps of 12". Courtesy Paris Observatory.*

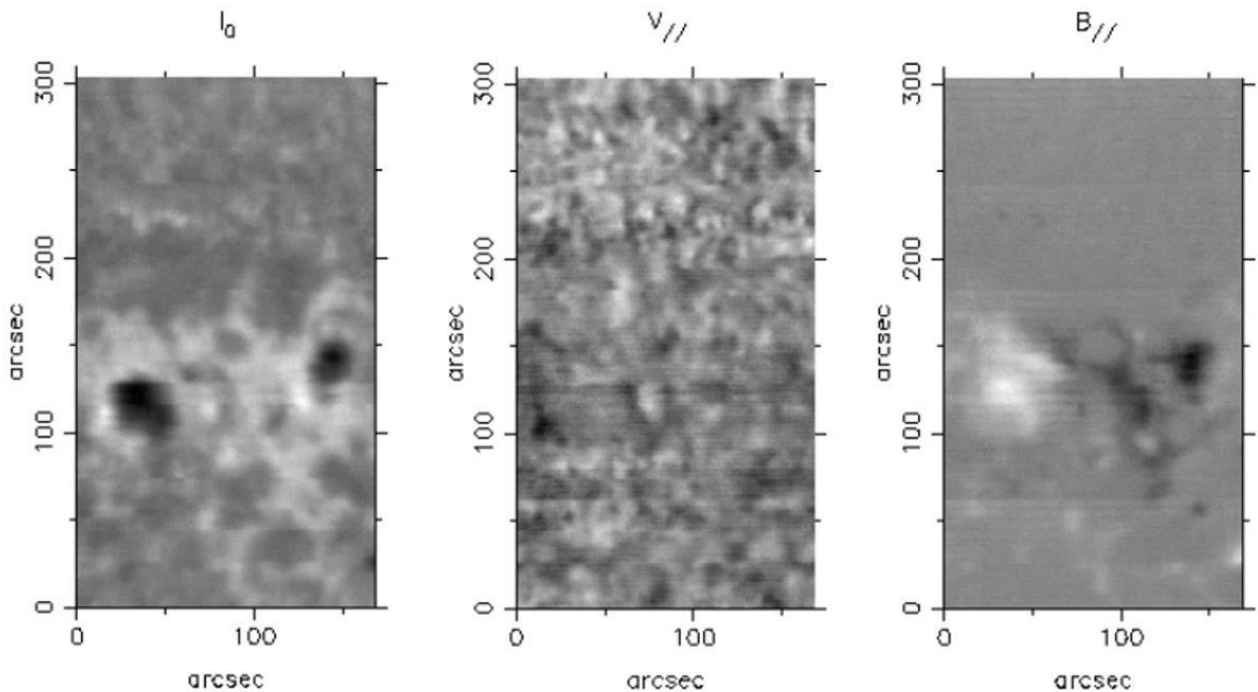

*Figure 17 : observation of an active region with the MSDP in Mg b2 5173 line in poor seeing conditions. The slicer and the polarimeter deliver 2 x 18 sub-channels for two states of circular polarization (I+V, I-V). The active region required a surface scan with 25 steps of 12" in 3 minutes, the elementary FOV being 15" x 170". Courtesy Paris Observatory.*

### 5 – b – Simulation of FeI 6173 (Landé factor g* = 2.5)

Figure 18 shows what could be done with the well known FeI 6173 line observed by HMI/SDO and SO/PHI. This is a narrow photospheric line (0.085 Å FWIP) and we want to record a 0.16 Å bandwidth around the line centre. With this chosen bandwidth, the full FOV ($0 < x < x_m/2$) is available in x-direction for LOS velocities less than 1 km/s. It is reduced by a factor 2 in the case of velocities less than 3 km/s, but it could be larger if one selects a smaller bandwidth (but it is important to keep the inflexion points inside the waveband). The spectral resolution (32 mÅ) is close to the one provided by the SOT spectro-polarimeter onboard HINODE, and better than the one available with tunable filters onboard space missions.

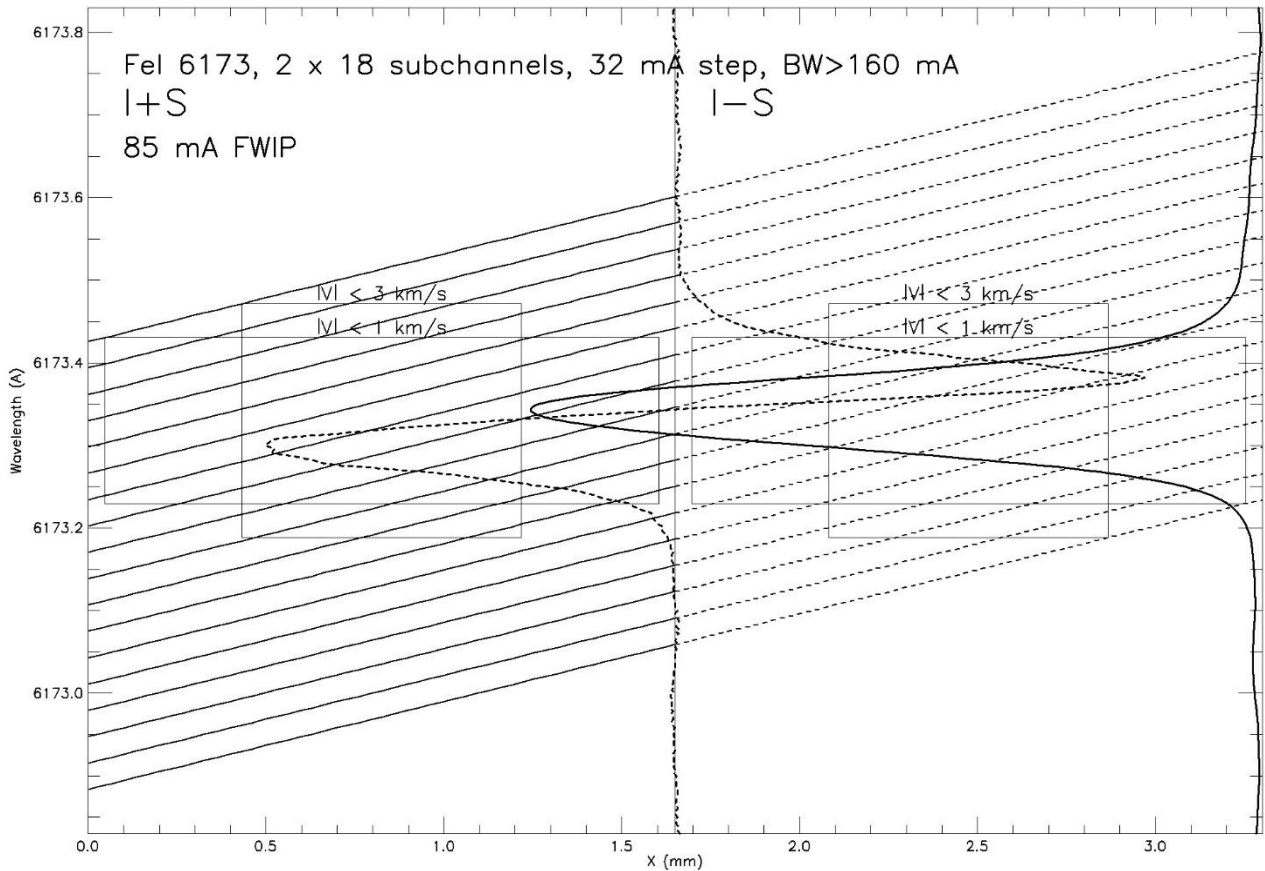

*Figure 18* : *wavelength functions of the 36 sub-channels for FeI 6173 (courtesy Paris Observatory).*

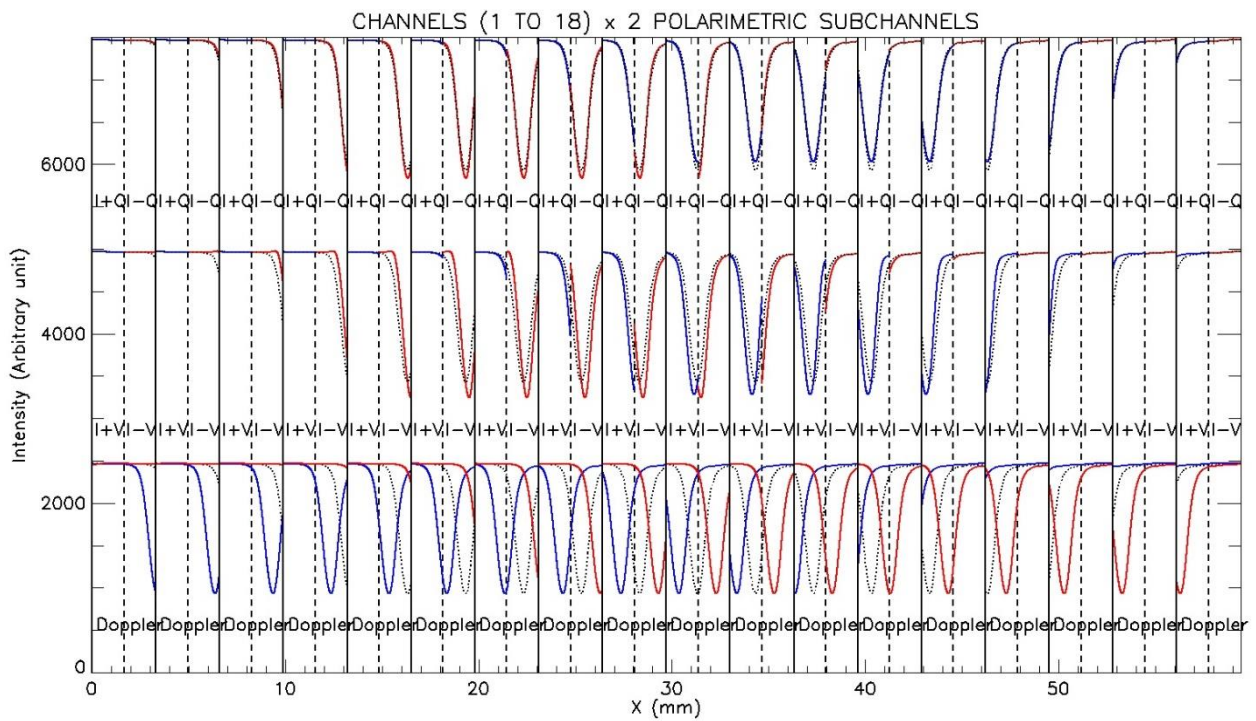

*Figure 19* : *FeI 6173 spectral line in the 2 x 18 sub-channels. Top: I+Q and I-Q. Centre: I+V and I-V. Both are computed for B = 500 G. Bottom: positive and negative dopplershifts (courtesy Paris Observatory).*

The sub-channels are strictly cospatial, but they do not have the same wavelength sampling, because, for a given abscissa of the FOV such that $0 < x < x_m/2$, there is a wavelength shift of $x_m/(2d)$. This phenomenon appears clearly in figures 19, 20 & 21 and implies interpolations to deliver Stokes combination profiles (such as I+S and I-S, S = Q, U, V) with the same wavelength sampling, and later derive I, Q and Q/I profiles.

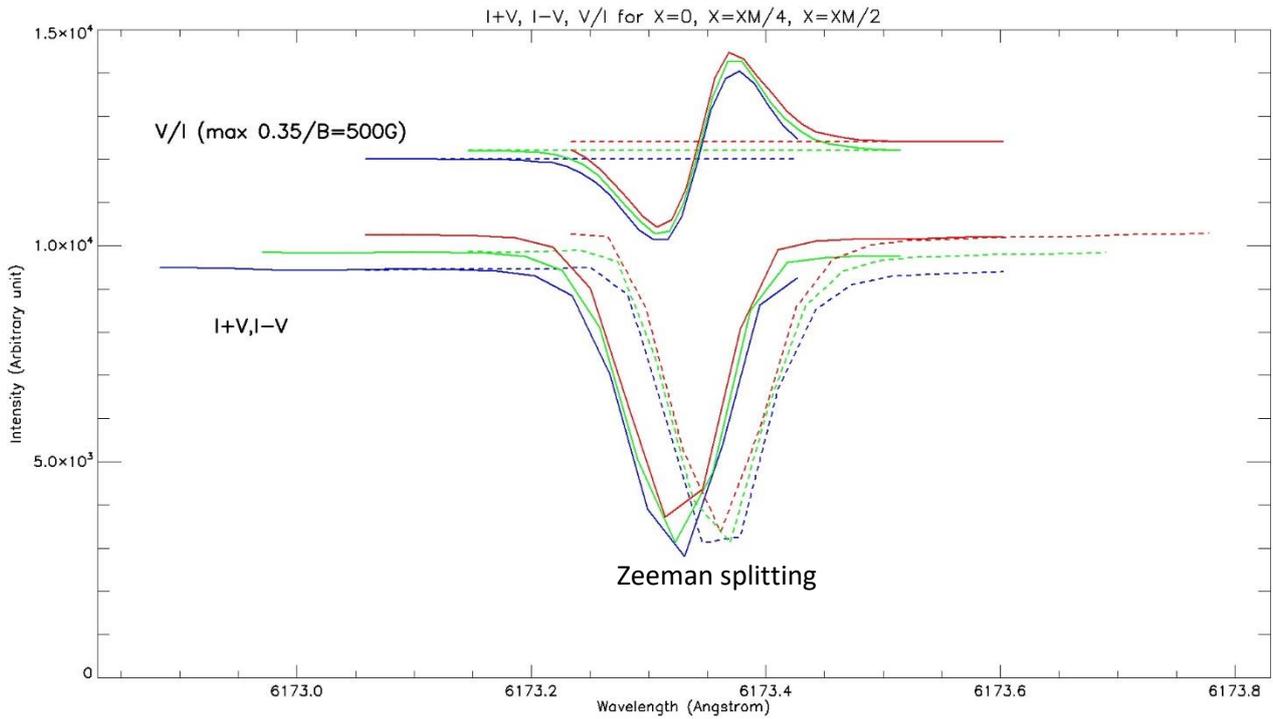

*Figure 20* : *simulation of I+V (solid), I-V (dashed) and V/I profiles of FeI 6173 spectral line for x = 0 (blue), $x_m/4$ (green) and $x_m/2$ (red). As the sampling wavelengths of I+V and I-V are not identical, an interpolation is necessary to get I, V and V/I. Here we took B// = 500 G (courtesy Paris Observatory).*

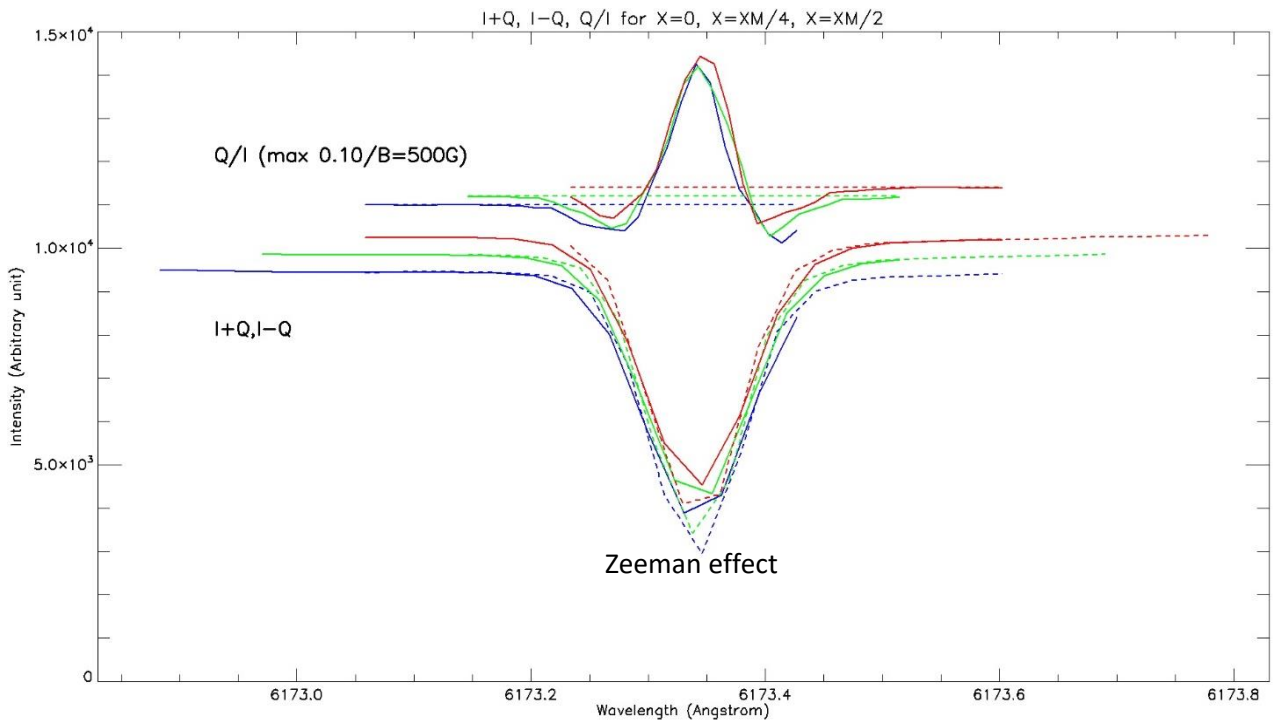

*Figure 21* : *simulation of I+Q (solid), I-Q (dashed) and Q/I profiles of FeI 6173 spectral line for x = 0 (blue), $x_m/4$ (green) and $x_m/2$ (red). As the sampling wavelengths of I+Q and I-Q are not identical, an interpolation is necessary to get I, Q and Q/I. We assumed B⊥ = 500 G (courtesy Paris Observatory).*

## 6 - On line data, processing methods, software guide

Observations with the MSDP, from 1977 to 2000, were recorded on 70 mm films which are archived at Meudon and can be digitized upon request. A CCD 1536 x 1024 was implemented in the early 2000s, and some CCD observations are available on-line at the BASS2000 solar archive:

https://bass2000.obspm.fr/longterm_archive.php?instrume=dpsm

The data processing software requires the free Gfortran (GNU project), and the latest version is available upon request to pierre.mein@obspm.fr

A detailed guide is available (link above) in two parts, the first one describes the methods used for the data reduction, and the second part is dedicated to the software and parameter significance.

A full MSDP bibliography is also available (link above).